\documentclass[aps,amssymb,prl, amsmath, twocolumn, floatfix]{revtex4}
\usepackage{graphics}
\usepackage{psfrag}
\usepackage{epsfig}
\usepackage{subfigure}
\begin{document}
\bibliographystyle{prsty}

\title{The Breakdown of Kinetic Theory in Granular Shear Flows}
\author{Gregg Lois$^{(1)}$}
\author{Ana\"el Lema\^{\i}tre$^{(1,2)}$}
\author{Jean M. Carlson$^{(1)}$}
\affiliation{
$^{(1)}$ Department of physics, University of California, Santa Barbara, California 93106, U.S.A.}
%\affiliation{$^{(2)}$ LMDH - Universite Paris VI, UMR 7603, 4 place Jussieu - case 86, 75005  Paris - France}
\affiliation{$^{(2)}$ Institut Navier-- LMSGC, 2 all\'ee K\'epler,77420 Champs-sur-Marne, France}
\date{\today}

\begin{abstract}
We examine two basic assumptions of kinetic theory-- binary collisions and molecular chaos-- using numerical simulations of sheared granular materials.  We investigate a wide range of densities and restitution coefficients and
 demonstrate that kinetic theory breaks down at large density and small restitution coefficients.  In the regimes where kinetic theory fails, there is an associated emergence of clusters of spatially correlated grains. 
\end{abstract}
\maketitle

For granular materials, kinetic theory has been the primary strategy used to systematically derive 
hydrodynamics equations, starting from elementary assumptions about grain-grain 
interactions~\cite{garzodufty}. 
This has led to much interest in applying predictions from the theory to
realistic granular flows 
(for reviews, see~\cite{reviews}), and recent work continues in this direction~\cite{compare,lutsko}.
However, kinetic theory strictly applies only to dilute gases, and the extent that it applies to the dense
regime remains unclear.

\begin{figure}[htbp]
\begin{center}
\vspace{-0.2 in}
\psfrag{yl}{\Huge{$1-e$}}
\psfrag{xl}{\Huge{$\nu$}}
\subfigure{\scalebox{0.37}{\includegraphics{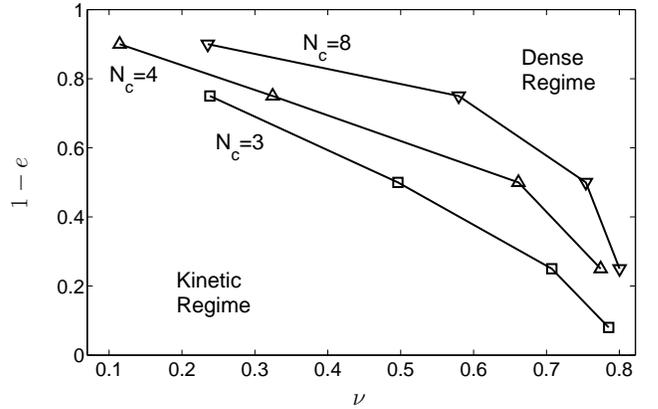}}}
\vspace{-0.2 in}
\caption{\label{bndry} Contours of the average cluster size $N_c$ as a function of restitution coefficient $e$ and packing fraction $\nu$.  Kinetic theory applies to binary interactions between grains ($N_c=2$).  }
\end{center}
\end{figure}

In this Letter we perform tests of the fundamental assumptions of kinetic theory, 
using the Contact Dynamics (CD) algorithm.   
We find that kinetic theory is severely limited by the assumption that only binary 
interactions occur between grains.
Instead, an effective many body interaction arises that
is a direct consequence of persistent contacts in the dense regime.
In Fig.~\ref{bndry} we characterize the failure of the binary collision assumption, using
spatial force correlations to approximate the average number of grains $N_c$ 
that form a cluster in contact.  As we see, the cluster size increases when going to low restitution coefficient and high density, which should limit the relevance of kinetic theory.
%The condition $N_c < 3$ 
%estimates (and perhaps overestimates~\footnote[-5]{The boundary should occur at $N_c=2$, corresponding to a binary collision, but noise in our numerics prevents such a precise determination (see Fig.~\ref{forcecorrs}).}) the regime where kinetic theory applies.  
This Letter provides quantitative estimates of this breakdown.  

Most kinetic theory research starts with the Boltzmann equation, which is derived from
the BBGKY hierarchy, 
and then finds its solutions~\cite{garzodufty, lutsko, savagejeffrey}.
However, certain assumptions are necessary to derive the Boltzmann equation. 
We begin by discussing two of these assumptions.

Consider a system of $N$ grains and the evolution equation for 
the $N$-body probability distribution function (pdf) $f^{(N)}({\bf\vec r},{\bf\vec p})$, 
where $({\bf\vec r},{\bf\vec p})$ 
denotes the set of all positions and momenta for the system 
(with the notation ${\bf\vec r}=\{\vec r_i\}$).
This equation is simply a statement of conservation of probability and reads
\begin{equation}
\frac{\partial f^{(N)}}{\partial t}
+\sum_{i}\,\frac{\partial f^{(N)}}{\partial{\vec r}_i} \cdot \frac{{\vec p}_i}{m}
=-\sum_{ij}\,\frac{\partial f^{(N)}}{\partial{\vec p}_i} \cdot {\vec F_{ij}}
\label{eqn:proba}
\end{equation}
where we have decomposed the force on each grain as a sum over pairs: $\vec F_i=\sum_j \vec F_{ij}$.
%~\footnote{It is an easy check that Eq.~(\ref{eqn:proba}) is equivalent to Liouville equation for a Hamiltonian system.}.

The BBGKY hierarchy is derived from Eq.~(\ref{eqn:proba}) by integration~\cite{Cercignani}.  This hierarchy is the set of 
$N-1$ equations for the evolution of the $n$-body pdfs
$f^{(n)}$, with $1 \leq n < N$. 
If the force resulting from a pair interaction depends only
on the positions and velocities of the interacting pair, 
%this hierarchy takes a simple form. 
%To determine the equation of motion for the function $f^{(n)}$
%we integrate out $N-n$ degrees of freedom in Eq.~(\ref{eqn:proba}).  Because interactions
%only occur between pairs of grains, this
%leaves in the rhs an integral of the function $f^{(n+1)}$. 
then the evolution equation of each $f^{(n)}$ depends only on $f^{(n+1)}$.  This
is the classical form of the BBGKY hierarchy.

\begin{figure}[htbp]
\begin{center}
\vspace{-0.15 in}
\mbox{
\subfigure{\scalebox{.1}{\includegraphics{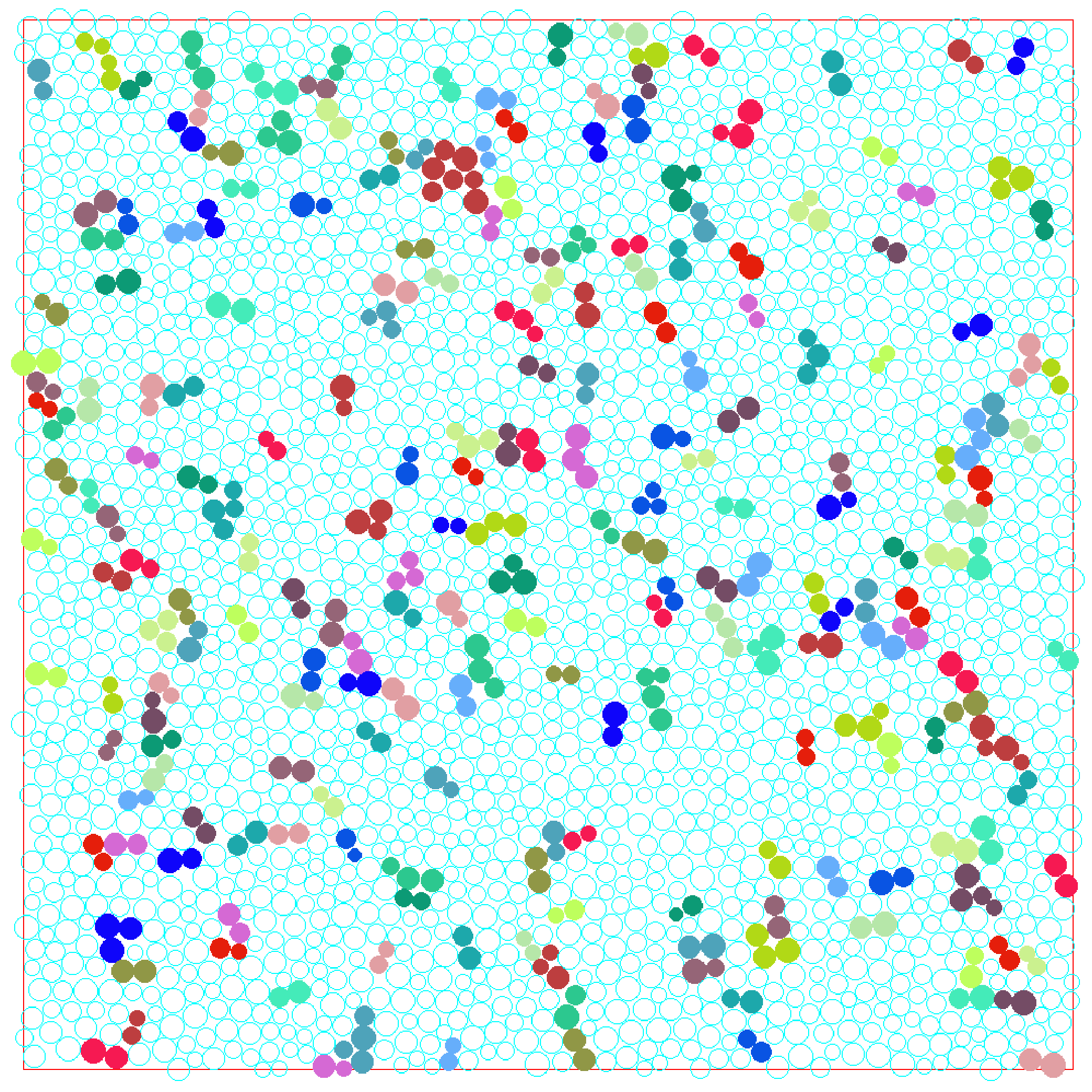}}}
}
\mbox{
\subfigure{\scalebox{.1}{\includegraphics{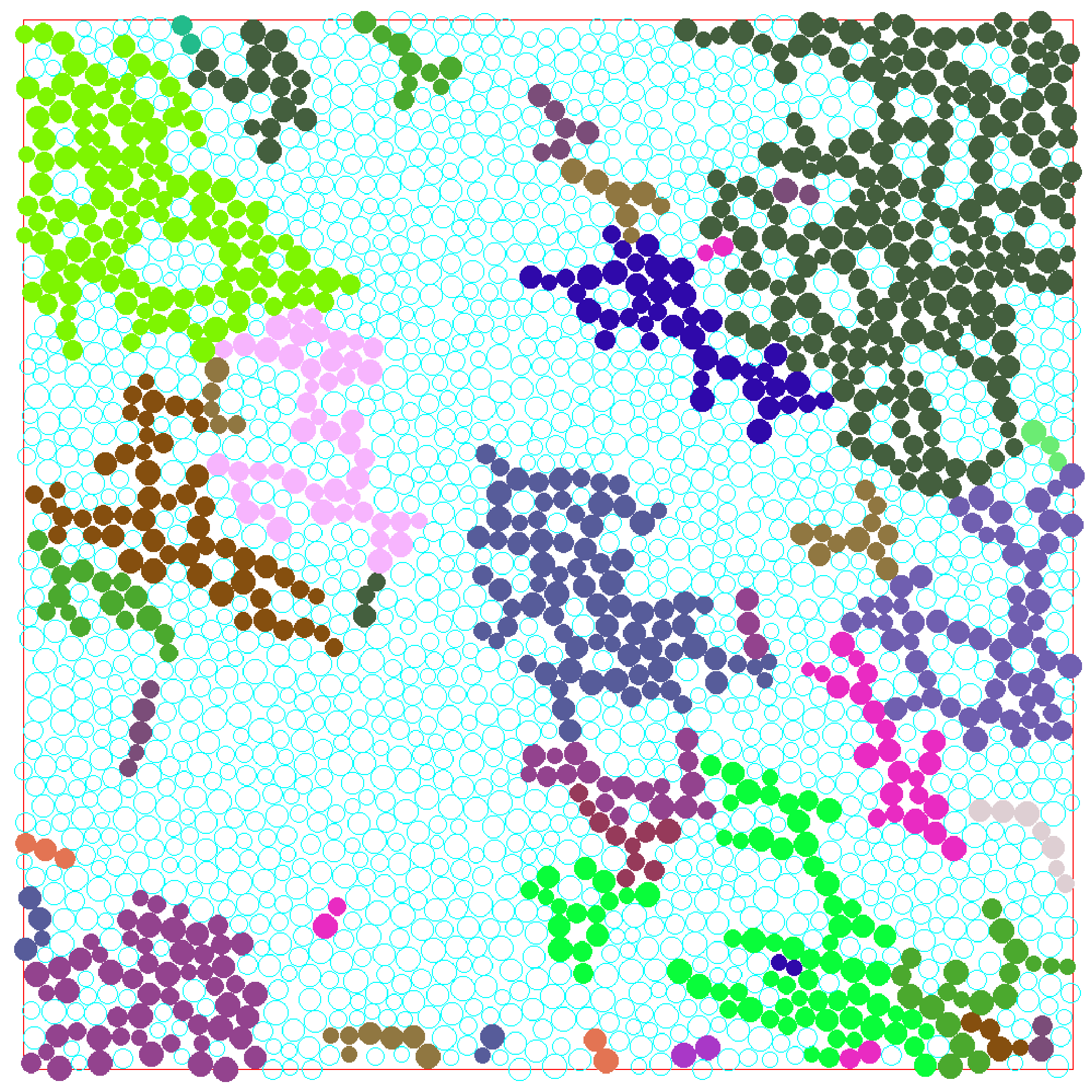}}}
}
\vspace{-0.15 in}
\caption{\label{e92e0} (Color online).  Steady state screenshots of sheared granular materials at two different restitution coefficients: $e=0.92$ (left) and $e=0$ (right).  Grains involved in a collision during a small time period are colored, with different colors denoting different contact networks.  Interactions are binary for $e=0.92$ whereas large clusters form for $e=0$.}
\end{center}
\end{figure}

\vspace{-0.15 in}
When applying the derivation of the BBGKY hierarchy to granular materials, additional care must be taken.
For dilute hard-sphere gases the derivation applies since
interactions result from binary collisions and thus pair forces
depend only on the positions and velocities of the 
interacting pair.
However, in dense systems of dissipative grains the situation is different.
%, and the 
%forces cannot be given any a priori expression which would only involve
%the position and velocities of the interacting pair. 
In this case there may 
be clusters of grains in persistent contact, as illustrated in Fig.~\ref{e92e0}.
If clusters have formed, the force between any pair in the cluster will depend {\em not} only on the positions and velocities
of the interacting pair, but also on the positions and velocities of all other grains in the cluster.
In this case, the hierarchical structure of the BBGKY equations is not guaranteed.  Therefore, in order to derive the 
BBGKY hierarchy for granular materials, we must make the {\em binary collision assumption}.  This stipulates
that only binary collisions occur, thereby assuring that pair forces depend only on the 
positions and velocities of the interacting pair. 

The Boltzmann equation follows from the first equation ($n=1$) of the BBGKY hierarchy, which relates $f^{(1)}$ to $f^{(2)}$. 
A second assumption is also required, the {\em molecular chaos assumption}, which simplifies this equation by setting
\begin{equation}
f^{(2)}(r_1,r_2,v_1,v_2) = \chi(r_1,r_2) f^{(1)}(r_1,v_1) f^{(1)}(r_2,v_2),
\label{molecularchaos}
\end{equation}
where $\chi$ describes possible correlations in the positions of particles.
This assumption allows us to approximate the first equation of the BBGKY hierarchy as a non-linear equation for $f^{(1)}$:  this is the Boltzmann equation.

The Boltzmann equation relies on the assumptions of binary collisions and molecular chaos.  In the rest of this Letter we test these fundamental assumptions,
using two-dimensional CD simulations~\cite{CDalgo} 
of frictionless granular materials in simple shear flow at constant volume.  
The simulations are performed using Lees-Edwards boundary conditions, which ensure translational invariance.  The density, restitution coefficient, and shear rate are prescribed and other observables are measured.  
%For all measurements presented here, we have chosen the time step small enough so that the results do not depend on its value.

In Fig.~\ref{e92e0} we show two representative screenshots from our simulations, 
in steady state, for identical shear rate and density, but different restitution 
coefficients, $e=0.92$ and $e=0$. A small time interval is chosen, and
in both cases grains that collide during this time interval are colored.  Different
colors corresponding to separate contact networks.  
For $e=0.92$ the interacting grains are well spaced and tend to occur in pairs, whereas
for $e=0$ the interacting grains tend to form large clusters. 
These clusters indicate the emergence of persistent contacts for small values of $e$ (high inelasticity).

\vspace{-0.15 in}
We now test the binary collision assumption, which is primarily a statement about the forces
between pairs of grains. The idea is to measure the relative contribution to momentum
transport from binary collisions and from clusters in persistent contact.
A quantitative measure of these contributions can be obtained by measures of the stress tensor.
The ``static'' stress ${\bf \Sigma}^s$ reads:
\begin{equation}
{\bf \Sigma}^s_{\alpha \beta} A = \frac{1}{2} \sum_{i>j} (D_i+D_j) {\bf \hat{n}}_{{\bf ij}\alpha} {\bf F}_{{\bf ij}\beta}
\label{static}
\end{equation}
where $\alpha$, $\beta$ denote components and $i$, $j$ denote grains,
$D_i$ is the diameter of grain $i$, ${\bf \hat{n}_{ij}}$ the unit normal vector 
at contact between the pair $(i,j)$, and $A$ is the area of the simulation cell.
This quantity measures the true momentum transport in a microcanonical configuration.
At each time step, the CD algorithm determines the contact forces ${\bf F}_{{\bf ij}}$ by upholding constraints 
relevant to perfectly rigid contact between grains~\cite{gaj}.

When a binary collision occurs, the final relative velocity of the colliding pair is set equal to the initial 
relative velocity, multiplied by $-e$. 
The CD algorithm calculates the force between the pair
based on the instantaneous impulse that produces this final relative velocity.
Because of the time-discretization, this ``binary collision force'' is approximated
over a time interval $\Delta t$ by a constant force equal to the instantaneous impulse divided by $\Delta t$:
${\bf F_{ij}}^{bc}=\frac{\mu_{ij}}{\Delta t}\,(1+e)\,({\bf v_i}-{\bf v_j}) \cdot \hat{\bf n}_{{\bf ij}}\,\hat{\bf n}_{{\bf ij}}$,
where $\mu_{ij}$ is the reduced mass of grains $i$ and $j$, 
and ${\bf v_i}$ is the pre-collisional velocity of grain $i$.  
However, when multiparticle collisions occur, the total forces ${\bf F_{ij}}$ differ from the binary collision forces ${\bf F_{ij}}^{bc}$.
Replacing ${\bf F_{ij}}$ by ${\bf F_{ij}}^{bc}$ in Eq.~(\ref{static}) provides the flux of momentum
that would be transported if all forces resulted from binary collisions:
\vspace{-0.05 in}
\begin{equation}      
{\bf \Sigma}^{bc}_{\alpha \beta} A =
\frac{1+e}{2 \Delta t} \sum_{i>j} \mu_{ij} (D_i+D_j) {\bf \hat{n}}_{{\bf ij}\alpha} {\bf \hat{n}}_{{\bf ij}\beta}  ({\bf v_i} - {\bf v_j}) \cdot {\bf \hat{n}_{ij}}.
\label{collisional}
\end{equation}    
We call this tensor the ``collisional'' stress tensor: it is defined at any time, even in the presence of 
multi-contact interactions, and is an approximation to the static stress. 

A theory that assumes binary collisions and is capable of taking into account all correlations and providing an exact
expression for the distribution of velocities between incoming pairs of grains would only
account for ${\bf \Sigma}^{bc}_{\alpha \beta}$, 
but never ${\bf \Sigma}^{s}_{\alpha \beta}$. Because most kinetic theories assume binary collisions, the core question is whether
${\bf \Sigma}^{bc}_{\alpha \beta}$ is a reasonable approximation 
to ${\bf \Sigma}^{s}_{\alpha \beta}$.

\begin{figure}[t]
\begin{center}
\mbox{
\psfrag{tl}{\Huge{$p^s/p^{bc}$}}
\scalebox{0.35}{\includegraphics{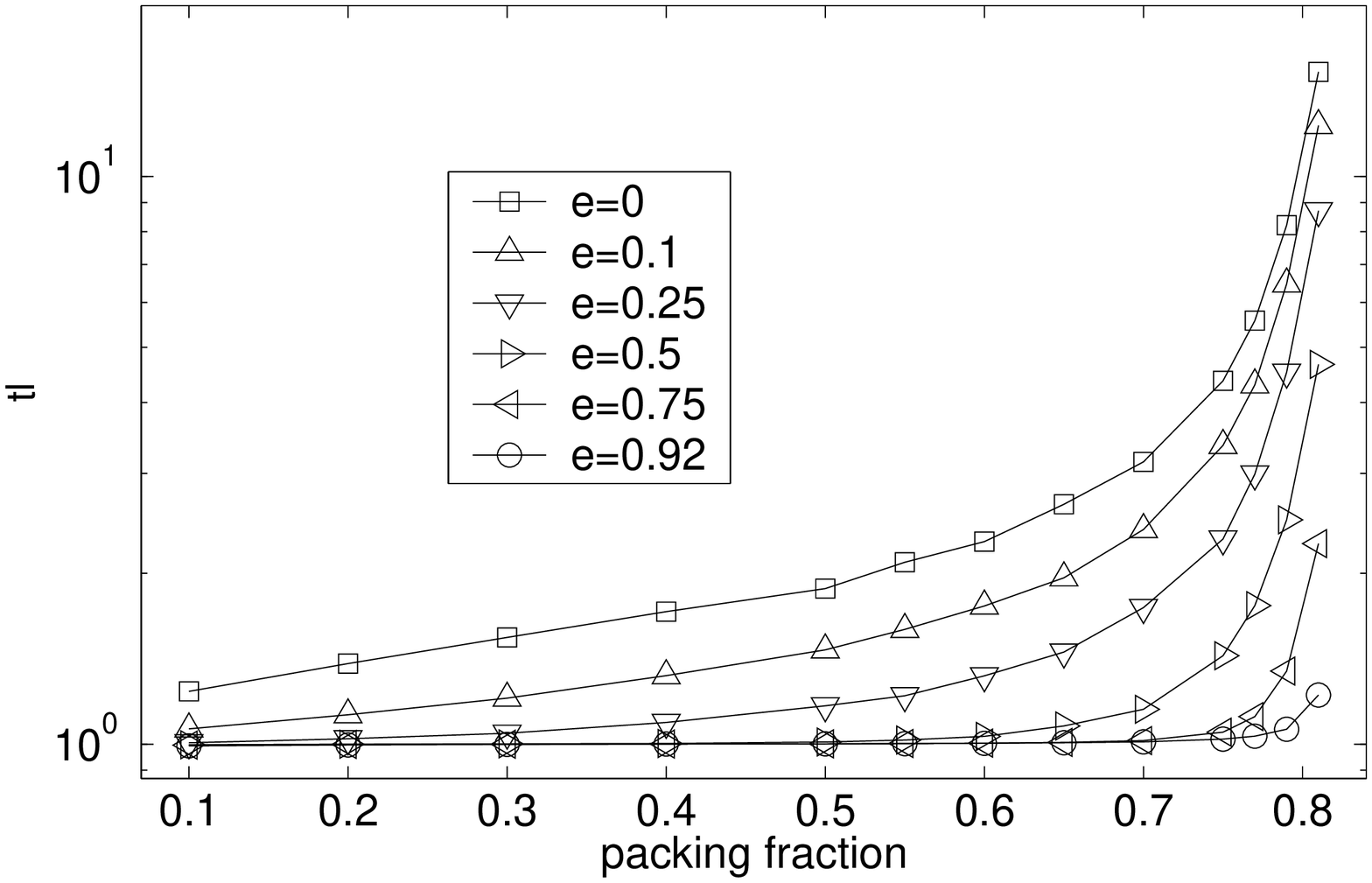}}
}
\mbox{
\psfrag{tl}{\Huge{$s^s/s^{bc}$}}
\scalebox{0.35}{\includegraphics{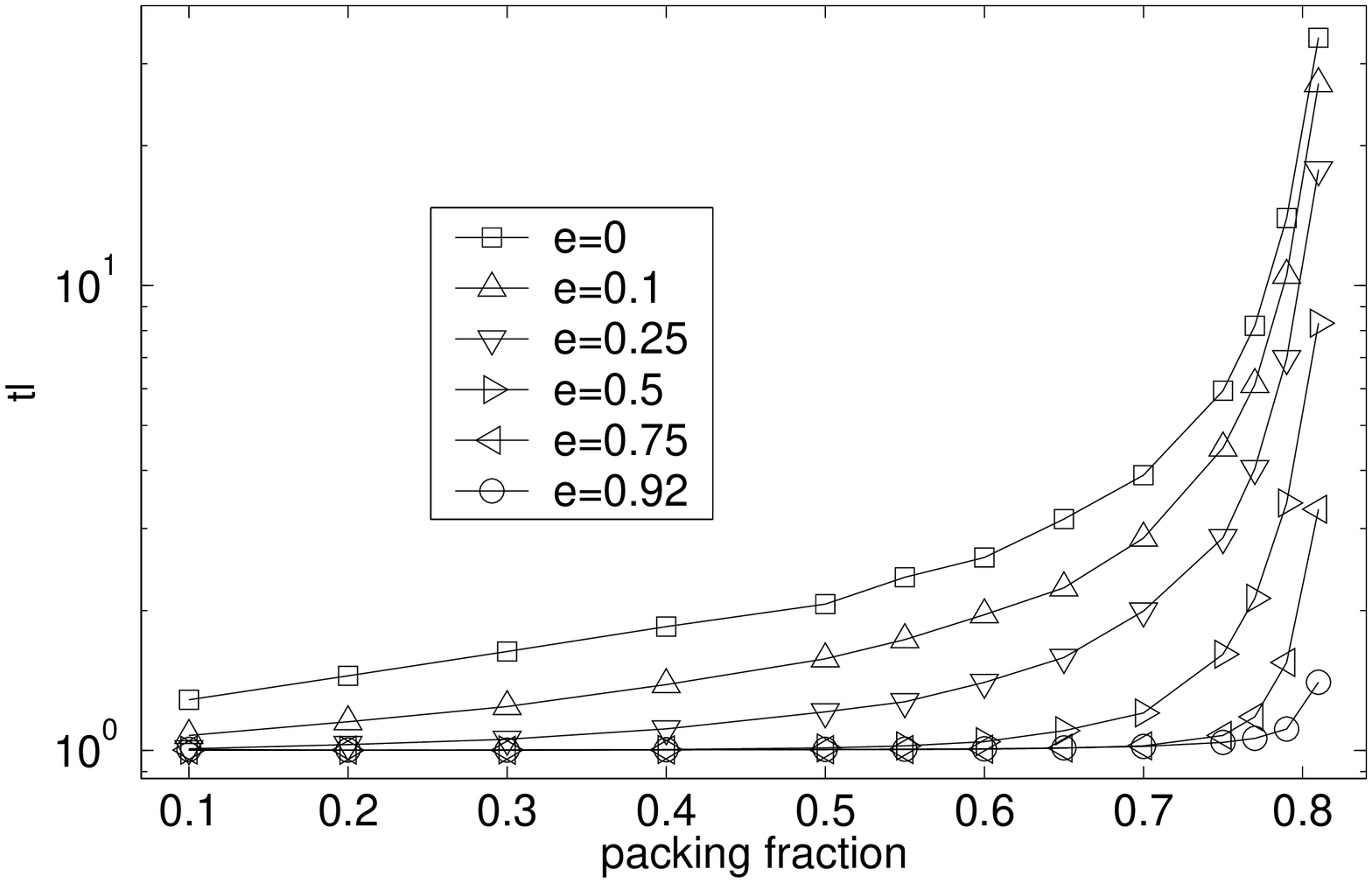}}
}
\vspace{-0.1 in}
\caption{\label{pressurestress} The static pressure divided by the collisional pressure ($p^s/p^{bc}$) and the static shear stress divided by the collisional shear stress ($s^s/s^{bc}$) as a function of packing fraction, for a variety of restitution coefficients $e$.  Values larger than one represent the breakdown of the binary collision assumption.  }
\end{center}
\end{figure}

To answer this question, we further decompose stresses into pressure $p$ and shear stress $s$.  
Pressure is one-half of the trace of the tensor and the shear stress is defined 
as either of the off-diagonal elements of the symmetric stress tensor.
In Fig.~\ref{pressurestress} we plot data from our simulations for the static pressure divided by the collisional pressure $p^s/p^{bc}$ and the static shear stress divided by the collisional shear stress $s^s/s^{bc}$ as a function of packing fraction, for a variety of restitution coefficients.  For restitution coefficients near unity and relatively low packing fraction, the static values are equal to the collisional values and the ratios in Fig.~\ref{pressurestress} are close to unity.  However, for large packing fractions and small restitution coefficients, the static values become larger than the collisional values.  This signals the breakdown of the binary collision assumption.  

This first numerical test quantitatively demonstrates that the collisional stress tensor is not an adequate approximation of the true static stress tensor in certain regimes of granular shear flow.  It rules out the possibility that a theory based on the binary collision assumption can be applied to predicting the static stress at high 
density or low restitution.

\vspace*{-0.33 in}
Next, we examine the
molecular 
chaos 
assumption: we test whether
a broad array of kinetic theories succeed in accounting for the collisional stress.
In two dimensions, kinetic theories that assume both binary collisions and molecular chaos
make a prediction for the collisional pressure $p^{kt}$~\cite{garzodufty, lutsko, savagejeffrey}:
\vspace{-0.05 in}
\begin{equation}
p^{kt} = (1+e) \chi \nu \, p^*
\label{kineticprediction}
\end{equation}
where $\nu$ is the packing fraction and $\chi$ is the pair correlation function at contact. 
This prediction is proportional to $p^*=n m \, \delta v^2/2$ where
$n$ is the number density, $m$ is the average mass, 
and $\delta v^2$ is the average square of the fluctuating velocity (the granular temperature).

Our second numerical test compares the prediction for the collisional pressure
to the actual collisional pressure measured in the simulations.
We determine all parameters in Eq.~(\ref{kineticprediction}) directly from the simulations:  
$e$ and $\nu$ are prescribed, $\chi$ and $p^*$ are measured.
Following other studies~\cite{lusa} we measure $\chi$ using the collision
frequency $\omega$ and the following formula from kinetic theory:
\vspace{-0.05 in}
\begin{equation}
\omega = \sqrt{2 \pi \delta v^2} \chi n \sigma
\label{kineticfrequency}
\end{equation}
where $\sigma$ is the average grain diameter.

\begin{figure}[t]
\begin{center}
\mbox{
\psfrag{yil}{\LARGE{$\langle \delta v_1 \delta v_2 \rangle /\delta v^2$}}
\scalebox{0.37}{\includegraphics{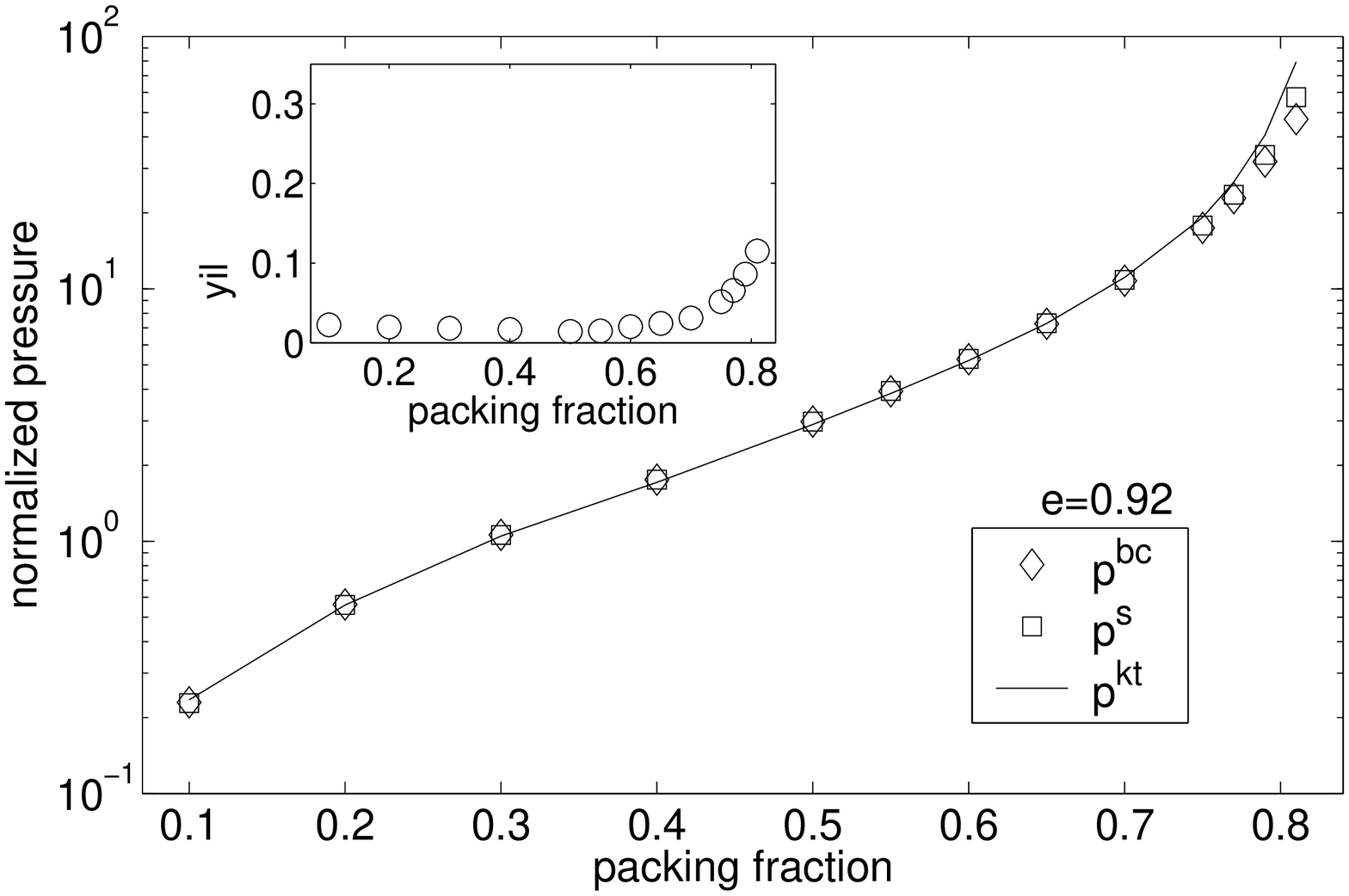}}
}
\mbox{
\psfrag{yil}{\LARGE{$\langle \delta v_1 \delta v_2 \rangle /\delta v^2$}}
\scalebox{0.37}{\includegraphics{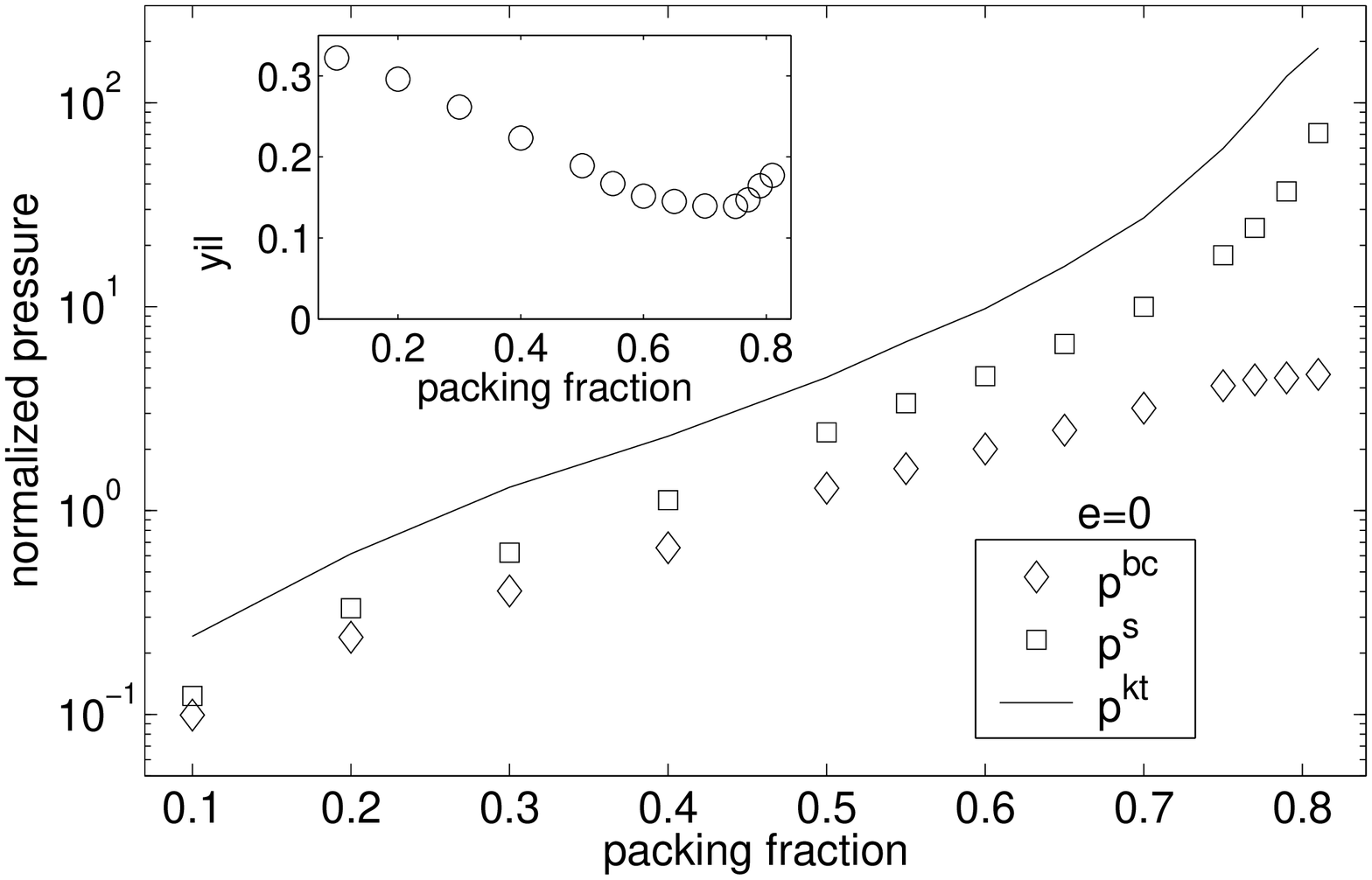}}
}
\vspace{-0.1 in}
\caption{\label{testenskog} Main figures:  The static pressure $p^s$, collisional pressure $p^{bc}$, and kinetic theory prediction $p^{kt}$, all normalized by $p^*$, for $e=0.92$ (top) and $e=0$ (bottom).  Insets:  Velocity correlations at contact, normalized by the granular temperature $\delta v^2$.  $p^{kt}$ from Eq.~(\ref{kineticprediction}) overestimates $p^{bc}$ when velocity correlations are large and positive.  }
\end{center}
\end{figure}

Using Eqs.~(\ref{kineticprediction}) and~(\ref{kineticfrequency}) 
we measure, without any fitting parameters, the approximation to the collisional 
pressure
resulting from the molecular chaos assumption. 
This is 
\\
\vspace*{-0.35 in}
\newline
reported in solid lines 
on Fig.~\ref{testenskog}, where we have also plotted raw data for both the collisional and static pressure. 
For $e=0.92$ there is excellent agreement between the 
kinetic theory prediction and the collisional pressure, even for large values of packing fraction.  
For $e=0$ the molecular chaos assumption leads to an overestimate of pressure 
at all packing fractions. 

We expect this overestimate to result from correlations
of the pre-collisional velocities:
if the velocities of two incoming grains are positively
correlated then their relative velocity is smaller, and the collisional pressure is thereby
reduced. Because the molecular chaos assumption does not incorporate these correlations, it overestimates
the collisional pressure.  

The insets of Fig.~\ref{testenskog}
contain measurements of the pre-collisional velocity correlations 
$\langle \delta v_1 \delta v_2 \rangle$, where $\delta v$ is the fluctuating part of the velocity 
in the direction parallel to the vector connecting the grain centers ${\bf \hat{n}_{ij}}$.
We measure correlations in this direction because only these velocity components
contribute to the collisional stress in Eq.~(\ref{collisional}).  
The average is performed over a disk centered on one grain with a radius of $1.8 \sigma$, 
although the results do not depend on the 
size of the averaging
disk.  
Pairs of
\\
\vspace*{-0.38 in}
\newline
grains that collided in the previous time step are excluded from the average in order 
to ensure the correlations are truly pre-collisional.  
We observe that large values of the correlation
correspond to packing fractions where kinetic theories based on the molecular chaos assumption
overestimate the collisional pressure.
However, the value of the correlation does not correspond to whether the 
collisional stress is a good approximation to the static stress-- this
seems to be related only to the breakdown of the binary collision assumption.

\begin{figure}[t]
\begin{center}
\psfrag{yl}{\Huge{$N_c$}}
\psfrag{yil}{\LARGE{$|C(\ell)|$}}
\psfrag{r}{\LARGE{$\ell$}}
\scalebox{0.36}{\includegraphics{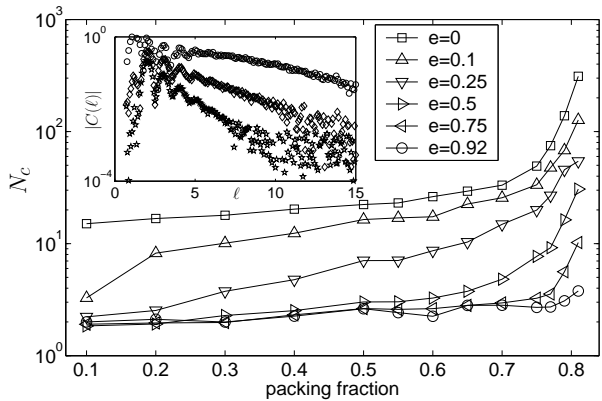}}
\vspace{-.15 in}
\caption{\label{forcecorrs} Inset: Plots of the spatial force-force correlations $|C(\ell)|$ (in arbitrary units on a log scale) for e=0 and three packing fractions $\nu=0.7$ (bottom) $\nu=0.77$ (middle) and $\nu=0.81$ (top).  Main Figure:  The average cluster size $N_c$ as a function of packing fraction, for many values of the restitution coefficient $e$.}
\end{center}
\end{figure}

In order to further understand how the binary collision assumption breaks down, we return 
to the observation that dense granular flows organize into clusters of interacting grains, as illustrated in Fig.~\ref{e92e0}.
The failure of the binary collision assumption is related to the formation of these clusters and the fact that
the stress tensor is not determined solely by two particle interactions.

A cluster of grains may be defined as a region in the material 
over which forces are correlated.
To determine the average number of grains $N_c$ in these correlated clusters, we 
measure the spatial force-force correlations $C(\ell) \equiv \langle 
{\vec F}(0) \cdot {\vec F}(\ell) \rangle$, where $\ell$ is a positive distance measured in grain diameters and ${\vec F}$ 
is the {\it total} vector force acting on a grain.  If a grain is isolated, so that there are no forces acting on it, it is not included in this average.
We then define $N_c$ proportional to the square of the correlation length
\vspace{-0.05 in}
\begin{equation}
\sqrt{N_c} \propto \langle \ell \rangle = \frac{\int \ell C(\ell) d\ell}{\int C(\ell) d\ell}
\label{NCd}
\end{equation}
and normalize so $N_c=2$ for $e=0.92$ and low density.  We choose this normalization because we have observed (see Fig.~\ref{pressurestress})
that the binary collision assumption, which corresponds to $N_c$=2, is appropriate for dilute, nearly elastic granular materials.  

Our measurements of $N_c$ are presented in 
Fig.~\ref{forcecorrs}, along with measurements of $C(\ell)$.  
The force correlations fluctuate at small distance and exhibit an exponential decay 
at large distance.  The values of $\langle \ell \rangle$ determined from Eq.~(\ref{NCd}) match the exponential decay (when we plot $e^{-\ell/\langle \ell \rangle} $) for 
the densities and restitution coefficients we have investigated.  
We notice from Fig.~\ref{forcecorrs} that the divergence of $N_c$ close to jamming
nicely echoes the divergence of pressure and shear stress ratios in Fig.~\ref{pressurestress}. 
This confirms that the formation of clusters is directly related to the breakdown of the 
binary collision assumption.  

The measurement of $N_c$ allows us to partition the phase space of granular shear flow into regions where
kinetic theory applies ($N_c=2$) and regions where it does not ($N_c>2$).  
In Fig.~\ref{bndry} we plot contours of $N_c$ as a function of restitution coefficient 
and packing fraction.
Although numerical noise prevents us from plotting the contour $N_c=2$, Fig.~\ref{bndry} 
provides an estimate of the regime where kinetic theory applies.
%-- for $N_c>3$, we 
%would not expect kinetic theory to hold.
%, where we have chosen a value larger
%than two to account for numerical noise in the measurement of $N_c$ at small values.         
%Shaded regions in Fig.~\ref{bndry} correspond to $N_c<3$:  they provides an estimate of regimes where kinetic theory applies.  Unshaded regions correspond
%to $N_c>3$:  this is the dense regime where kinetic theory does not apply.

We have presented two numerical tests of the fundamental assumptions of kinetic theory in granular 
materials:  first we have observed the breakdown of the binary collision assumption for large densities
and small restitution coefficients; second we have demonstrated that the molecular chaos assumption
is not valid for small restitution coefficients, due to pre-collisional velocity correlations.  In order 
for an approach based on kinetic theory to be useful at high density, the deficiencies in these core assumptions must be addressed. 
Although the molecular chaos assumption can in principle be addressed by incorporating velocity dependent terms in 
Eq.~(\ref{molecularchaos}), it seems to us that the failure of the binary collision assumption may be much more
difficult to overcome in a standard kinetic theory.  Successful theories of granular materials in the dense regime must incorporate clustering.

This work was supported by the W. M. Keck Foundation, the MRSEC program of the NSF under Award No. DMR00-80034, the J. S. McDonnell Foundation, NSF Grant No. DMR-9813752, the Lucile Packard Foundation, the Mitsubishi Corp., and the NSF under Grant No. PHY99-07949.


\vspace{-0.2 in}
\begin{thebibliography}{100}
\vspace{-0.2 in}

\bibitem{garzodufty}
V. Garz\'{o} and J. W. Dufty, Phys. Rev. E {\bf 59}, 5895 (1999);

\bibitem{reviews}
C. S. Campbell, Annu. Rev. Fluid Mech. {\bf 22}, 57 (1990); I. Goldhirsch, Annu. Rev. Fluid Mech. {\bf 35}, 267 (2003).

\bibitem{compare}
J. M. Montanero, V. Garz\'{o}, M. Alam, and S. Luding, cond-mat/0411548 (2004);
N. Mitarai and H. Nakanishi, Phys. Rev. Lett. {\bf 94}, 128001 (2005).

\bibitem{lutsko}
J. F. Lutsko, Phys. Rev. E {\bf 70} 061101 (2004).

\bibitem{savagejeffrey}
C. K. K. Lun, S. B. Savage, D. J. Jeffrey, and N. Chepurniy, J. Fluid Mech. {\bf 140},223 (1984);
J. T. Jenkins and M. W. Richman, Phys. Fluids {\bf 28}, 3485 (1985).

\bibitem{Cercignani} 
Carlo Cercignani {\em Theory and Application of the Boltzmann Equation} (Scottish Academic, London, 1975).

\bibitem{CDalgo}
J.~J. Moreau, Eur. J. Mech. A-Solids {\bf 13}, 93 (1994);
M. Jean, Comput. Methods Appl. Mech. Engrg. {\bf 177}, 235 (1999).
\bibitem{gaj}
G. Lois, A. Lema\^{\i}tre, and J. M. Carlson, cond-mat/0501535 (2005)

\bibitem{lusa}
S. Luding and A. Santos, J. Chem. Phys. {\bf 121}, 8458 (2004).

\end{thebibliography}
\end{document}